\documentclass{article}
\usepackage{ijcai25}

\usepackage{times}
\usepackage{soul}
\usepackage{url}
\usepackage[hidelinks]{hyperref}
\usepackage[utf8]{inputenc}
\usepackage[small]{caption}
\usepackage{graphicx}
\usepackage{amsmath}
\usepackage{amsthm}
\usepackage{booktabs}
\usepackage{algorithm}
\usepackage{algorithmic}
\usepackage[switch]{lineno}
\usepackage{amssymb}
\usepackage{multirow}
\usepackage{extarrows}
\usepackage{newfloat}
\usepackage{listings}
\usepackage{xcolor}
\usepackage{subcaption}
\linenumbers

\title{Boosting Knowledge Graph-based Recommendations through Confidence-Aware
Augmentation with Large Language Models}

\author{
Rui Cai$^1$\and
Chao Wang$^{2*}$\and
Qianyi Cai$^4$\and
Dazhong Shen$^3$\and
Hui Xiong$^{4}$\footnote{corresponding authors: Chao Wang and Hui Xiong}\\
\affiliations
$^1$Xi'an Jiaotong University \and
$^2$University of Science and Technology of China \and
$^3$Nanjing University of Aeronautics and Astronautics \and
$^4$The Hong Kong University of Science and Technology (Guangzhou)\\
\emails
yansoncai007@gmail.com,
wangchaoai@ustc.edu.cn,
qcai603@connect.hkust-gz.edu.cn,
dazh.shen@gmail.com,
xionghui@ust.hk
}

\begin{document}
\nolinenumbers
\maketitle

\begin{abstract}
Knowledge Graph-based recommendations have gained significant attention due to their ability to leverage rich semantic relationships. However, constructing and maintaining Knowledge Graphs (KGs) is resource-intensive, and the accuracy of KGs can suffer from noisy, outdated, or irrelevant triplets. Recent advancements in Large Language Models (LLMs) offer a promising way to improve the quality and relevance of KGs for recommendation tasks. Despite this, integrating LLMs into KG-based systems presents challenges, such as efficiently augmenting KGs, addressing hallucinations, and developing effective joint learning methods. In this paper, we propose the Confidence-aware KG-based Recommendation Framework with LLM Augmentation (CKG-LLMA), a novel framework that combines KGs and LLMs for recommendation task. The framework includes: (1) an LLM-based subgraph augmenter for enriching KGs with high-quality information, (2) a confidence-aware message propagation mechanism to filter noisy triplets, and (3) a dual-view contrastive learning method to integrate user-item interactions and KG data. Additionally, we employ a confidence-aware explanation generation process to guide LLMs in producing realistic explanations for recommendations. Finally, extensive experiments demonstrate the effectiveness of CKG-LLMA across multiple public datasets.
\end{abstract}

\section{Introduction}
Recommender systems~\cite{setrank,afdgcf} have become essential tools for content creation and personalization, especially in addressing the information overload issue. A promising approach within this domain is the integration of Knowledge Graphs (KGs)~\cite{kgat,kgrec,kgil}, which utilize KG triplets as side information to enrich user/item representations by encoding additional item-wise semantic relationships. This method effectively mitigates the data sparsity problem commonly encountered in traditional recommender systems~\cite{lightgcn,sgl}. However, constructing and maintaining KGs is resource-intensive, requiring significant human and computational efforts. Moreover, ensuring the accuracy of the knowledge in KGs is fraught with challenges, such as the introduction of noise in the form of irrelevant or erroneous triplets~\cite{kgil}, and the static nature of many KGs~\cite{kgsurvey} limits the ability to revise the outdated relations. These limitations highlight the need for more adaptable and less labor-intensive methods to maintain the relevance and robustness of KGs in recommender systems.

Traditional KG-based methods focus on extracting auxiliary content from KG for modeling user/item representations. For example, KGCN~\cite{kgcn} incorporates high-order entities' information in KGs, and KGAT~\cite{kgat} applies attentive aggregation for user-item-entity joint graph. More recently, research has shifted towards addressing the inherent noise within KGs, acknowledging that not all entities and relations are equally reliable. 
For example, KGCL~\cite{kgcl} performs contrastive learning to address noisy entities in KG, while KGRec~\cite{kgrec} drops triplets with lower rationale scores before building contrastive objectives. However, these methods are constrained by their reliance solely on information provided by the KG itself, which may be inherently noisy, outdated, or incomplete. 
Fortunately, recent advancements~\cite{llmrecsurvey} in Large Language Models (LLMs) have shown that LLMs demonstrate efficacy in refining KGs through eliminating spurious triplets and augmenting potential entity relationships~\cite{llmkgedit} by leveraging their sophisticated semantic comprehension capabilities and extensive knowledge repositories. 
Therefore, we aim to boost KG-based recommendations with LLM's assistance.

Notwithstanding the potential of LLMs to enhance KG-based recommender systems, several critical challenges remain unaddressed. Primarily, the token limit constraints inherent in existing LLMs necessitate a robust prompting design to augment all triplets in the whole KG effectively and consistently. 
Secondly, the susceptibility of LLMs to hallucination phenomena~\cite{llmhallucination} poses a significant risk of introducing erroneous triplets into original KGs. A robust confidence mechanism is essential to enhance the denoising process by accurately identifying potentially erroneous triplets within the context of recommendation tasks. 
Thirdly, the design of a unified learning framework that optimally leverages both the original KG and the newly added triplets to learn robust representations of users and items remains a challenge. 
Moreover, it is interesting and important to explore approaches to combine augmented KGs with LLMs for generating explanations for user behaviors, thereby enhancing the credibility of recommender systems.

In this paper, we introduce the Confidence-aware KG-based Recommendation Framework with LLM Augmentation (CKG-LLMA) to address the challenges in enhancing Knowledge Graph (KG) recommendations. Our framework begins with two-view LLM-based subgraph augmenter tailored for recommendation tasks to extract relevant knowledge within token constraints. To avoid LLM-induced noise and ensure the quality of downstream recommendation, we implement a Confidence-aware MOE Message Propagation mechanism, which adaptively retains or removes triplets, effectively integrating meaningful KG entity semantics into item representations with Mixture-of-Experts. Furthermore, to better leverage knowledge from both KG and LLM within an ID-based recommendation system perspective, we employ a Dual-view Two-step Contrastive Learning schema that links KG noise elimination with user-item interaction modeling, enhancing the robustness of user and item representations. 
Finally, the augmented KG and trained model are utilized to guide LLMs in generating realistic and informative explanations for recommendations, demonstrating the effectiveness of our approach. Extensive experiments on three public datasets indicate that our model outperforms various baseline methods and its ability to generate explainable and reliable recommendation results.

\vspace{-0.2cm}\section{Related Works}
Related works can be divided into two main categories: KG-based recommendations and unifying KGs and LLMs.

\subsection{KG-based Recommendations}
Existing KG-based approaches can be roughly classified into three categories: embedding-based, path-based, and GNN-based methods. 
Embedding-based techniques, such as CKE~\cite{cke}, enhance recommendation performance by integrating collaborative filtering with modeling of diverse item-related side information.
Path-based approaches like KPRN~\cite{KPRN} employ LSTM~\cite{lstm} networks to model relational meta-paths from KGs. However, the fixed design of these meta-paths limit their scalability and general applicability.
Currently, GNN-based methods are considered the forefront KG-based recommendation endeavors with remarkable efficiency. For example, KGAT~\cite{kgat} employs graph attention networks~\cite{gat} to prioritize neighbor aggregation based on their relative importance, while KGIN~\cite{kgin} enhances this approach by integrating user-specific relational embeddings into the aggregation process. Similarly, KGCL~\cite{kgcl} introduces KG semantics to mitigate data noise in recommendation systems through knowledge-guided contrastive learning, utilizing KG-aware data augmentation to probe the influence of items on user modeling. KGRec~\cite{kgrec} implements an knowledge rationalization mechanism that assesses triplets with rationale scores, subsequently integrating them into MAE-based \cite{mae} reconstruction learning and building contrastive objectives. However, these methods are limit in fixed KG, inhibiting their ability to capture and leverage the useful dynamic semantic information extracted by LLMs.

\vspace{-0.2cm}\subsection{Unifying KGs and LLMs} 
The integration of LLMs and KGs can be categorized into three primary approaches: KG-enhanced LLMs, LLM-augmented KGs, and synergistic combinations. 
KG-enhanced LLMs, exemplified by Think-on-graph~\cite{think-on-graph}, utilize KGs to guide LLMs in producing reasoned outputs. Conversely, LLM-augmented KGs leverage LLMs' knowledge editing capabilities to enhance KG performance in downstream tasks, as demonstrated by MPIKGC~\cite{mpikgc} which query LLMs to enriches context in knowledge graph completion task. 
Synergistic approaches, such as GreaseLM~\cite{GreaseLM}, aim to create frameworks where LLMs and KGs mutually enhance each other's capabilities. In the recommendation domain, methods like LKPNR~\cite{lkpnr} exploit LLMs' text processing abilities to improve content personalization. On the contrary, we propose to integrate LLMs and KGs bidirectionally, using LLMs to augment KG-based recommenders while leveraging the enriched KGs to guide LLMs in generating explanations.

\begin{figure*}[t]
\vspace{-0.4cm}
 \centering
  \includegraphics[width=0.9\textwidth]{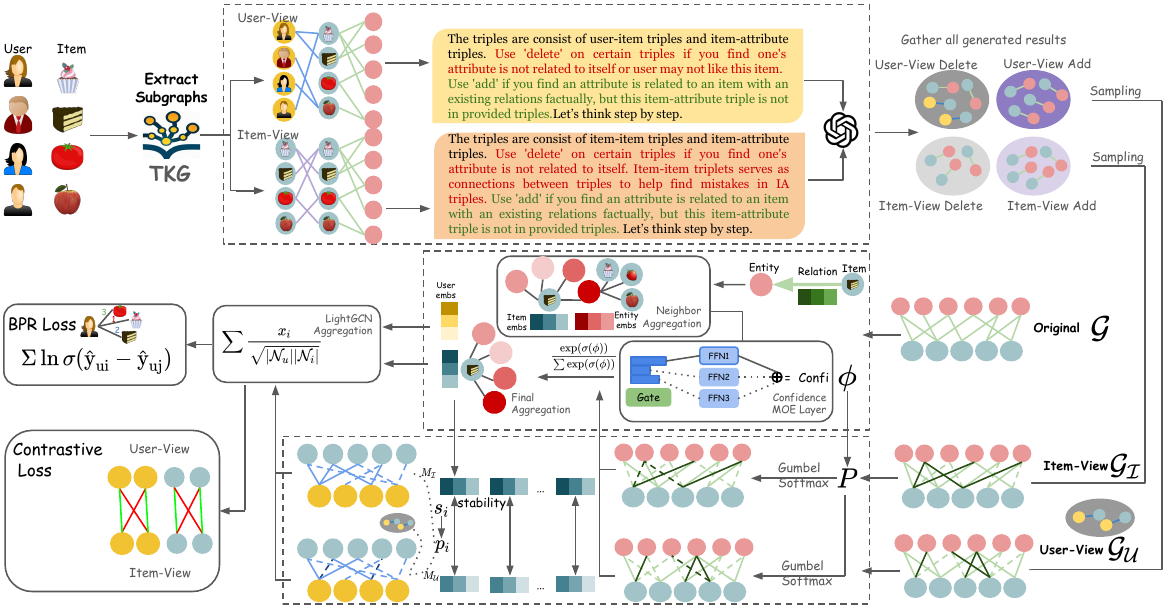}
  \caption{The overview of the proposed CKG-LLMA framework for knowledge graph based recommendations.}
  \vspace{-0.4cm}
  \label{fig:pipeline}
  \vspace{-0.2cm}
\end{figure*}

\vspace{-0.2cm} \section{Methodology}
We present the overall architecture of CKG-LLMA in Figure~\ref{fig:pipeline}. Details are discussed in following sub-sections.

\vspace{-0.2cm}\subsection{Definition and Overview}
We first introduce the notions and definitions
used throughout this paper, and and formally defining the LLM-enhanced KG-based recommendation task.

\noindent \textbf{User-Item Interaction Graph (IG)} In the typical recommendation scenario, there is a user set $\mathcal{U}$ and an item set $\mathcal{I}$. A bipartite graph $\mathcal{Y} \in |\mathcal{U}| \times |\mathcal{I}|$ is used to represent the collaborative signals among users and items, denoted as $\mathcal{Y} = \{(u, y_{ui}, i)|u \in \mathcal{U},i \in \mathcal{I} \}$, where $y_{ui}$ = 1 indicates an observed interaction of user $u$ over item $i$, and vise versa. 

\noindent \textbf{Tripartite Relational Knowledge Graph (TKG)} We define the concept of TKG as a heterogeneous graph contains 3 kinds of triplets.
\begin{itemize} \setlength{\itemsep}{0pt} \setlength{\leftmargini}{0pt}
\item \textit{UI triplets} User behaviors are triplets ($u, interact, i$) to represent the corresponding signal $y_{ui}$=1 for user $u$ and item $i$ in IG, which denoted as $\mathcal{G}_{UI} = \{(u, interact, i)| (u, 1, i) \in \mathcal{Y}\}$.
\item \textit{IA triplets} Item knowledge are side information for items (e.g., item attributes or external knowledge), and we denote IA triplets (e.g., (\textit{apple}, \textit{has category}, \textit{fruit})) records such triplet facts between item set $\mathcal{I}$ and entity set $\mathcal{A}$ as $\mathcal{G}_{IA} = \{(i, r, a)|i \in \mathcal{I}, r \in \mathcal{R}_{IA}, a \in \mathcal{A}\}$.
\item \textit{II triplets} Item correlations are auxiliary triplets that capture the shared relationships with entities between items. For instance, \textit{apple} and \textit{banana} both have category \textit{fruit}, so there exists a triplet (\textit{apple}, \textit{same category}, \textit{banana}) in II triplets. We define the set of item correlations as $\mathcal{G}_{II} = \{(i, r', j)| r' \in \mathcal{R}_{II}, i, j \in \mathcal{I}\}$. 
\end{itemize}
Then the three triplets set are unified to represent $\mathcal{G} = \mathcal{G}_{UI} \cup \mathcal{G}_{II}  \cup \mathcal{G}_{IA} = \{(h, r, t)|h,t\in \mathcal{E}, r \in \mathcal{R}\}$, where $\mathcal{E}=\mathcal{U} \cup \mathcal{I} \cup \mathcal{A}$ and $\mathcal{R}=\mathcal{R}_{IA} \cup \mathcal{R}_{II}\cup \{interact\}$.
Then, given IG graph $\mathcal{Y}$ and the unified knowledge graph $\mathcal{G}$, our goal is to develop a predictive function $\mathcal{F}(u, i|\mathcal{Y}, \mathcal{G}, \Theta)$ that estimates the probability of interaction between a user $u \in \mathcal{U}$ and an item $i \in \mathcal{I}$ with learnable parameters $\Theta$.


\noindent \textbf{Pipeline Overview} Our model CKG-LLMA first augmented the knowledge graph $\mathcal{G}$ 
by extracting subgraphs and building LLM prompts in two views through an LLM-based Subgraph augmenter, where the outputted augmentation advice can be classified into 4 pools with different purposes and views. 
Then we calculate the confidence of each triplet for dropout to diminish the latent noise within the origin KG and LLM-generated augmentation, and we design a Confidence-aware MOE Message Propagation methanism to effectively fuse the side information with items. 
We train our model through a joint learning approach with a BPR loss and a contrastive loss, while we sample triplets in pools and build contrastive objectives through a Two-step Dual-view process. 
After training, we generate the explanation of user-item interaction through LLM with the learned confidence of each triplet as assistance.

\vspace{-0.2cm}\subsection{LLM-based Subgraph Augmenter}
Manual knowledge graphs often have inaccuracies due to outdated or incorrect relationships, impacting recommendations. To enhance the quality of knowledge graphs, 
we propose to refine and augment the extracted knowledge subgraphs with the goal of knowledge completion by LLMs.


\vspace{-0.2cm}\subsubsection{Subgraph Extraction and Prompt}
Due to token limitations of LLMs, it is impractical to enhance using information from the entire graph. Instead, we utilize subgraph-based enhancements with LLMs. For $N$ collaborative singals $\mathcal{Y}^{sub} = \{(u_{k}, y_{u_{k}i_{k}}=1, i_{k})\}^{N}_{k=1}$, we extract all related triplets $\mathcal{G}^{sub}$ in all three background KG $\mathcal{G}_{IA}$, $\mathcal{G}_{II}$, and $\mathcal{G}_{UI}$, denoted as $\mathcal{G}^{sub} = \mathcal{G}_{IA}^{sub} \cup \mathcal{G}_{UI}^{sub} \cup \mathcal{G}_{II}^{sub}$. 
 
 Past work~\cite{LLMRec} demonstrates that LLMs process item side information as explicit knowledge, employing fact verification for triple augmentation. Conversely, LLMs treat user behaviors as implicit knowledge, augmenting these triples through reasoning. Based on this phenomenon, we design 2 kinds of extraction logic (User-view and item-view) with unique prompts to guide LLM in enhancing KG from a multi-view perspective.
\begin{itemize} \setlength{\itemsep}{0pt} \setlength{\leftmargini}{0pt}
    \item \textbf{User-view Extraction} The subgraph, denoted as $\mathcal{G}^{sub}_{\mathcal{U}}$, consists of $\mathcal{G}_{UI}^{sub}$ and $\mathcal{G}_{IA}^{sub}$. With an user-based prompt, we want LLM focuses on both rectifying explicit item-attribute triplets through fact verification, and pruning user-item interaction through implicit reasoning. 
    \item \textbf{Item-view Extraction} The subgraph, denoted as $\mathcal{G}^{sub}_{\mathcal{I}}$, consists of $\mathcal{G}_{IA}^{sub}$ and auxiliary triplets $\mathcal{G}_{II}^{sub}$. With an item-based prompt, we want LLM focuses on rectifying the item-attribute triplets with explicit knowledge and auxiliary triplets of item correlations. 
\end{itemize}

\vspace{-0.2cm}\subsubsection{LLM Augmentation ($\mathcal{P}^{add}_{\mathcal{U}}$, $\mathcal{P}^{del}_{\mathcal{U}}$, $\mathcal{P}^{add}_{\mathcal{I}}$, $\mathcal{P}^{del}_{\mathcal{I}}$)}
Our prompt design leverages the LLM's ability to detect contextual nuances in diverse subgraph patterns and provide modification advice enriched with domain-specific information.
We transform the ID-based triplets into textual information for LLM understanding using prompts. 
In user-view prompting, we expect LLM to rectify $\mathcal{G}_{IA}^{sub} \in \mathcal{G}^{sub}_{\mathcal{U}}$ with addition and deletion, and prune $\mathcal{G}_{UI}^{sub} \in \mathcal{G}^{sub}_{\mathcal{U}}$ with deletion. 
The add and delete triplets are saved as two pools, denoted as \textbf{$\mathcal{P}^{add}_{\mathcal{U}}$} and \textbf{$\mathcal{P}^{del}_{\mathcal{U}}$}, while $\mathcal{P}^{del}_{\mathcal{U}} = \mathcal{P}^{del}_{\mathcal{U}_{UI}} \cup \mathcal{P}^{del}_{\mathcal{U}_{IA}}$ involves 2 sets representing the fact verification and reasoning results. In item-view prompting, we expect LLM to thoroughly rectify $\mathcal{G}_{IA}^{sub} \in \mathcal{G}^{sub}_{\mathcal{I}}$ and respectively save two pools of modificationa as \textbf{$\mathcal{P}^{add}_{\mathcal{I}}$} and \textbf{$\mathcal{P}^{del}_{\mathcal{I}}$}. 
Details can be found at Appendix.

\vspace{-0.2cm}\subsection{Confidence-aware MOE Message Propogation}
Due to the existance of hallucination problems in LLM,  
erroneous generated triplets in pools $\mathcal{P}^{add}_{\mathcal{U}}$ and $\mathcal{P}^{add}_{\mathcal{I}}$ could bring meaningless links in origin $\mathcal{G}$ as new noise.
To reduce such potential disturbances and effectively aggregate knowledge within KG structures, we introduce a learnable edge confidence for each item-attribute triplet, considering the relation heterogeneity. This approach assigns a confidence score for LLM-enhanced edges, significantly boosting their robustness in handling diverse relational data.

To enhance the correlations between attributes and items and better incorporate with LLM introduced information, inspired by GAT~\cite{gat}, we initially apply an aggregation layer on items of $\mathcal{G}$, where the updated item embedding $\mathbf{x}_i'$ is computed as:
{\small\begin{equation}
\label{equ:gat-fuse}
    \!\!\!\!\!\!\mathbf{x}_i'\! =\! \mathbf{x}_i \!+ \!
    \sum_{k=1}^{|\mathcal{N}_{i}|}\frac{exp(\sigma\!(\mathbf{a}_1^\mathrm{T}\left[\mathbf{W}_1 \mathbf{x}_{i} || \mathbf{W}_1 \mathbf{x}_{a_k} \right]))}{\sum_{j=1}^{|\mathcal{N}_{i}|}exp(\sigma(\mathbf{a}_1^\mathrm{T}\left[\mathbf{W}_1 \mathbf{x}_{i} || \mathbf{W}_1\mathbf{x}_{a_j} \right]))}\mathbf{x}_{a_k} \text{,}\!\!\!
\end{equation}}
where $\mathbf{W}_1 \in \mathbb{R}^{d \times d}, \mathbf{a}_1 \in \mathbb{R}^{d}$ are attention weights in each layer and $\sigma$ is the \textit{LeakyReLU} activation function.

Then we introduce LLM generated triplets to $\mathcal{G}$ with an designed confidence algorithm. Firstly, to reflect the contribution of each triplet for user preference, a confidence weighting function $\phi(\cdot)$ is implemented with a designed confidence-aware mixture of experts~\cite{moe} layer $\xi(\cdot)$. Specifically, for triplet $t=(i, r, a)$, its confidence is: 
{\small\begin{equation}
\label{equ:confidence}
    \phi(i, r, a) = \xi(\mathbf{x}_r, \mathbf{W}\left[\mathbf{x}_i'||\mathbf{x}_a\right]),
\end{equation}}
where $\mathbf{x}_i$, $\mathbf{x}_r$ and $\mathbf{x}_a \in \mathbb{R}^{d}$ are the embedding of $i$, $r$ and $a$. $\mathbf{W} \in \mathbb{R}^{d \times 2d}$ is a learnable transition matrix. $\xi(\cdot)$ contains a set of feed-forward neural networks $\{E_i(x)\}_i^{N_e}$ of $N_e$ experts. 
In order to capture the fine-grained information behind KG triplets, $\xi(\cdot)$ first computes the gate value of each expert $i$ based on entity features $\mathbf{f_t} = \mathbf{W}\left[\mathbf{x}_i'||\mathbf{x}_a\right]$ of $t$ as:
{\small\begin{equation}
\label{equ:moe1}
    \omega_i = \frac{e^{\mathbf{W_r} \mathbf{f_t}}_i}{\Sigma_{j}^{N_e} e^{\mathbf{W_r} \mathbf{f_t}}_j},
\end{equation}}
then the output computation of $\xi$ is the linearly weighted combination of each expert’s computation as:
{\small\begin{equation}
\label{equ:moe2}
    \xi(\mathbf{x}_r, \mathbf{f_t}) = \Sigma_i^{N_e} \omega_i \mathbf{x}_r^T\textit{E}_i(\mathbf{f_t}),
\end{equation}}
where we use linear layers for each expert $E$ here. 
We further calculate each triplet's confidence with renewed embeddings $\mathbf{x}_i'$ in equation \ref{equ:confidence}. After obtaining the confidence $\phi(t)$ for each triplet $t=(i, r, a) \in \mathcal{G}$, we aggregate the confidence-aware attribute embeddings to generate final aggregated item representations $\mathbf{x}_i''$ as:
{\small\begin{equation}
\label{equ:confi-fuse}
    \mathbf{x}_i'' = \mathbf{x}_i' + 
    \sum_{k=1}^{|\mathcal{N}_{i}|}\frac{exp(\sigma(\phi(i, r, a)))}{\sum_{j=1}^{|\mathcal{N}_{i}|}exp(\sigma(\phi(i, r_j, a_j)))}\mathbf{x}_{a_k}.
\end{equation}}

\vspace{-0.2cm}\subsection{Dual-view Two-step Contrastive Learning}
While aggregating side information to enhance item representations, in order to learn more essential aspects (e.g., extracting useful information for the recommendation task from the KGs and reducing the impact of noise data in the KGs) in user preferences modeling from the KG refined by large language models, we construct a dual-view contrastive learning framework, as the inherent connections between external elements can be utilized to direct data augmentation through cross-view self-supervised signals. We first adopt a differentiable KG augmentation strategy, then enhance IG from a "View-and-Stability" perspective. 

\vspace{-0.2cm}\subsubsection{Step1: Differentiable Knowledge Graph Augmentation}
To incorporate the LLM-generated information (User-view $\&$ Item-view) into origin $\mathcal{G}$, we first randomly sample triplets from $\mathcal{P}^{add}_{\mathcal{U}}$ and $\mathcal{P}^{add}_{\mathcal{I}}$ at a ratio of $\mu_a$ relative to the number of triplets in $\mathcal{G}$, and similarily sample triplets from $\mathcal{P}^{del}_{\mathcal{U}}$ and $\mathcal{P}^{del}_{\mathcal{I}}$ in a ratio $\mu_d$ to form 4 sets as:
{\small\begin{equation}
\label{equ:sampletriplets}
    \begin{aligned}
    & M_{\mathcal{U}}^{add} = \{m_{i}^{\mathcal{U}add}\}_{i=1}^{N_{add}}, 
    M_{\mathcal{U}}^{del} = \{m_{i}^{\mathcal{U}del}\}_{i=1}^{N_{del}}, \\
    & M_{\mathcal{I}}^{add} = \{m_{i}^{\mathcal{I}add}\}_{i=1}^{N_{add}},
    M_{\mathcal{I}}^{del} = \{m_{i}^{\mathcal{I}del}\}_{i=1}^{N_{del}},
    \end{aligned}
\end{equation}}
During the knowledge aggregation process, we only consider combining information from IA triplets $\mathcal{G}_{IA}$ as this is the main modified part in KG augmentation. We incorporate above sampled triples into $\mathcal{G}$ and build two different KGs $\mathcal{G}_{\mathcal{U}}$ and $\mathcal{G}_{\mathcal{I}}$ from two views as initially augmented graphs, i.e., 
{\small\begin{equation}
    \begin{split}
        \mathcal{G}_{\mathcal{U}} = \mathcal{G} \cup M_{\mathcal{U}}^{add} \setminus M_{\mathcal{U}_{IA}}^{del}, ~~~\mathcal{G}_{\mathcal{I}} = \mathcal{G} \cup M_{\mathcal{I}}^{add} \setminus M_{\mathcal{I}}^{del}.
    \end{split}
\end{equation}}

To reduce potential noise within the origin $\mathcal{G}$ and brought by LLM generated information, we apply the confidence for each triplet in graphs to drop links by learnable probabilities. Specifically, we calculate the confidence of each triplet $t = (i, r, a)$ in $\mathcal{G}_{\mathcal{U}}$ and $\mathcal{G}_{\mathcal{I}}$ as $\phi(t)$ with Equation \ref{equ:confidence}. We then transform the confidence $\phi(t)$ into a selection probability $P(t)$ using function $P(\cdot)$ as :
{\small\begin{equation}
\label{equ:dropprob}
    P(t) = Sigmoid(\phi(t) * K),
\end{equation}}
where $K$ is a scalable hyper-parameter to control the probability. This is the kept probability $P(t)$ for each triplet $t$ after the information incorporation of KG and LLM. Following~\cite{gumbelsoftmax}, gumbel softmax strategy is used to approximate the discrete sampling with differential consistency, allowing gradient propagation in this augmentation process.
With sampling decision $d_t$, we gather all decisions in two graphs $\mathcal{G}_{\mathcal{U}}$ and $\mathcal{G}_{\mathcal{I}}$ to renew two graphs by an operator $\psi(\cdot)$:
{\small\begin{equation}
\label{equ:gumbelrenew}
    \psi(\mathcal{G}_{\mathcal{U}}) = \mathcal{G}_{\mathcal{U}} \odot D_\mathcal{U}, \psi(\mathcal{G}_{\mathcal{I}}) = \mathcal{G}_{\mathcal{I}} \odot D_\mathcal{I},
\end{equation}}
where $D_\mathcal{U} = \{d_t|t \in \mathcal{G}_{\mathcal{U}}\}$ and $D_\mathcal{I} = \{d_t|t \in \mathcal{G}_{\mathcal{I}}\}$ are the sets of selection decisions for triples in augmented KGs.

\vspace{-0.0cm}\subsubsection{Step2: IG Enhancement via View-and-Stability}
To fully exploit the LLM-enhanced information, we augment the interaction graph combining views and a designed item stability. After the knowledge graph augmentation procedure, we derive $ \psi(\mathcal{G}_{\mathcal{U}})$ and $ \psi(\mathcal{G}_{\mathcal{I}})$ as view-specific augmented KGs and consolidate the graphs with item embeddings through Equations \ref{equ:gat-fuse} and \ref{equ:confi-fuse} and obtain view-specific item representations $\mathbf{x}''_{\mathcal{U}}$ and $\mathbf{x}''_{\mathcal{I}}$. Similar as~\cite{kgcl}, to explore the invariance property of item representations in different augmented views, we define the cross-view stability $s_i$ of item $i$ between the representations aggregated from knowledge of distinctive views though cosine similarity function $C(\cdot)$ of embeddings as:
{\small\begin{equation}
\label{equ:stability}
    s_i = C(\mathbf{x}''_{{\mathcal{U}}_{i}}, \mathbf{x}''_{{\mathcal{I}}_{i}}).
\end{equation}}
The cross-view stability score $s_i$ serves as an indicator of an item's resilience to changes in topological information. Items with higher $s_i$ values demonstrate less sensitivity to such changes. 
The stability indicator provides a way to mitigate the negative effects of both knowledge graph dependencies and noise in IG $\mathcal{Y}$ by incorporating auxiliary self-supervised signals, thereby guiding the IG augmentation. 
Specifically, we tend to drop item nodes in $\mathcal{Y}$ with lower stability, so we adopt a similar process to transform stability into keeping the probability of items as:
{\small\begin{equation}
\label{equ:stability2}
    \nonumber p_i = \frac{1-p_{drop}}{mean_i(exp(s_i))} * \frac{exp(s_i) - max_i(exp(s_i))}{max_i(exp(s_i)) - min_i(exp(s_i))},
\end{equation}}
where $p_{drop}$ is a preset inherent drop probability. 
With the keep probability $p_i$ of item $i$, we further generate two masking vectors $\mathbf{M}_{\mathcal{U}}$, $\mathbf{M}_{\mathcal{I}} \in \{0, 1\}$ based on the Bernoulli distribution~\cite{bernoulli}. Then we apply these two masks onto user-item interaction graph $\mathcal{Y}$ with the sampled deletion set $M_{\mathcal{U}_{UI}}^{del}$ from early LLM generated information at user view as:
{\small\begin{equation}
\label{equ:stabmasking}
\begin{aligned}
        & \zeta(\mathcal{Y})_{\mathcal{U}} = \{(u, y_{ui}, i) | u \in \mathcal{U}, i \in \mathbf{M}_{\mathcal{U}} \odot \mathcal{I}\} \setminus \mathcal{Y}_{del}, \\
        & \zeta(\mathcal{Y})_{\mathcal{I}} =  \{(u, y_{ui}, i) | u \in \mathcal{U}, i \in \mathbf{M}_{\mathcal{I}} \odot \mathcal{I}\} ,
\end{aligned}
\end{equation}}
where $\mathcal{Y}_{del} = \{(u, y_{ui}, i)|(u, interact, i) \in M_{\mathcal{U}_{UI}}^{del}\}$ is the elaborately sampled and then filtered user-item pairs from User-view delete pool $\mathcal{P}^{del}_{\mathcal{U}}$. 

\vspace{-0.2cm}\subsubsection{Confidence-aware Contrastive Learning}
Once we obtain the two view-specific augmented IG $\zeta(\mathcal{Y})_{\mathcal{U}}$ and $\zeta(\mathcal{Y})_{\mathcal{I}}$ from a dual augmentation process, we adopt a LightGCN-like ~\cite{lightgcn} message propagation strategy to further refine the knowledge aggregated representations and encode the collaborative effects from user-item interactions as:
{\small\begin{equation}
\label{equ:lightgcn}
    \nonumber \mathbf{x}_u^{(l+1)} = \sum_{i \in \mathcal{N}_u}\frac{\mathbf{x}_i^{(l)}}{\sqrt{|\mathcal{N}_u||\mathcal{N}_i|}}, \mathbf{x}_i^{(l+1)} = \sum_{u \in \mathcal{N}_i}\frac{\mathbf{x}_u^{(l)}}{\sqrt{|\mathcal{N}_i||\mathcal{N}_u|}},
\end{equation}}
where $\mathbf{x}_u^{(l)}$ and $\mathbf{x}_i^{(l)}$ are the propogated representations of user $u$ and item $i$ in $l$-th graph layer. $\mathcal{N}_u$ and $\mathcal{N}_i$ denotes the neighbor nodes of $u$ and $i$ in IG, which are $u$'s interacted items and $i$'s connected users respectively. We totally set $L$ layers, and the initial embeddings of items serves as input $\mathbf{x}_i^{0}$ of first propogation layer are the confidence-aware knowledge aggregated embeddings $\mathbf{x}''_{i}$. We obtain the final user/item embeddings by summing up the representations from all $L$ layers of LightGCN and taking the average as:
{\small\begin{equation}
\label{equ:lightgcn-average}
    \mathbf{x}_u = \frac{1}{L}\sum_{l=1}^{L}\mathbf{x}^{l}_u, \mathbf{x}_i = \frac{1}{L}\sum_{l=1}^{L}\mathbf{x}^{l}_i
\end{equation}}

To maximize the mutual information between augmented views and enhance the learnt embeddings' robustness, particularly focusing on confidence-aware aspects, we adopt constrastive learning from two step enhanced views, $\zeta(\mathcal{Y})_{\mathcal{U}}$ and $\zeta(\mathcal{Y})_{\mathcal{I}}$. After obtaining the final user/item embeddings from equation \ref{equ:lightgcn-average}, we have User-view embeddings ($\mathbf{x}^{\mathcal{U}}_{u}$,$\mathbf{x}^{\mathcal{U}}_{i}$) and Item-view embeddings ($\mathbf{x}^{\mathcal{I}}_{u}$,$\mathbf{x}^{\mathcal{I}}_{i}$). We divide the contrastive loss into 2 side: In user node side, users $u_1$ and $u_2$ constitute the positive pairs ($\mathbf{x}^{\mathcal{U}}_{u_1}$,$\mathbf{x}^{\mathcal{I}}_{u_1}$), ($\mathbf{x}^{\mathcal{U}}_{u_2}$,$\mathbf{x}^{\mathcal{I}}_{u_2}$) are generated from the same user of different views, and the negative pairs ($\mathbf{x}^{\mathcal{U}}_{u_1}$,$\mathbf{x}^{\mathcal{I}}_{u_2}$), ($\mathbf{x}^{\mathcal{U}}_{u_2}$,$\mathbf{x}^{\mathcal{I}}_{u_1}$) are generated from different users in both graph views. Similar in item node side. The general form of node contrastive loss is:
{\small\begin{equation}
\label{equ:conloss}
    \nonumber \mathcal{L}_{con} = \sum_{n \in \mathcal{V}} -\log \frac{\exp\left( C\left(\mathbf{x}_n^1, \mathbf{x}_n^2\right)/\tau \right)}{\sum_{n' \in \mathcal{V}, n' \neq n} \exp\left( C\left(\mathbf{x}_n^1, \mathbf{x}_{n'}^2\right)/\tau \right)},
\end{equation}}
where the subscript $n$ represents either user or item. $\mathcal{V} = \mathcal{U} \cup \mathcal{I}$ and $\tau$ is a temperature parameter. $C(\cdot)$ is the cosine similarity function.

\vspace{-0.2cm}\subsection{Model Optimization}
For the main recommendation task, we use the dot product between user \& item representations as the prediction, which is denoted as $\hat{y}_{ui} = \mathbf{x}_u^\mathrm{T}\mathbf{x}_i$. To optimize the model parameters, we adopt Bayesian Personalized Ranking loss to optimize thhe model parameters as follows:
{\small\begin{equation}
\label{equ:bprloss}
    \mathcal{L}_{bpr} = \sum_{(u, i, j)\in \mathcal{D}} -log \sigma(\hat{y}_{ui} - \hat{y}_{uj}),
\end{equation}}
We construct the training instances $\mathcal{D} = (u, i, j)$, where $(u, 1, i) \in \mathcal{Y}$ is the ground-truth and $(u, 0, j) \in \mathcal{Y}$ is a randomly sampled negative interaction. To optimize all loss functions, we use a joint learning approach with the following overall loss function:
{\small\begin{equation}
\label{equ:jointloss}
    \mathcal{L} = \mathcal{L}_{bpr} + \lambda_c * \mathcal{L}_{con} + \lambda_{\theta} *||\Theta||_2^2,
\end{equation}}
where $\lambda_c$ and $\lambda_{\theta}$ are loss weight parameters and $\Theta$ represents the learnable model parameters.

\vspace{-0.2cm}\subsection{Confidence-aware Explanation Generation}
With the trained predictive function $\mathcal{F}$ to forecast user-item interactions, we turn to construct an explanation generation module elucidating user behaviors. Specifically, we first combine two add pools $\mathcal{P}^{add}_{\mathcal{U}} \cup \mathcal{P}^{add}_{\mathcal{I}}$ and filter out triplets less than a threshold $\mu$ to construct an unified augmented KG as $\mathcal{G}^{aug}$. Then for user $u \in \mathcal{U}$ and any interacted item $i \in \mathcal{I}$, we aim to generate a natural language sentence to justify why $i$ is recommended to $u$. We design a two-step chain-of-thought~\cite{cot} reasoning procedure to instruct LLM for this explanation generation task:

\textbf{KG Information Extraction} We first extract domain-specific information for user $u$ and item $i$ from the TKG $\mathcal{G}$: a) UI triples: We extract all interactions between user $u$ and its interacted items $\mathcal{I}_{u} = \{j|(u, 1, j) \in \mathcal{Y}\}$ to form $\mathcal{G}'_{UI} = \{(u, interact, j)|j \in \mathcal{I}_{u}\}$. b) II triples: We find all items in $\mathcal{I}_{u}$ that contain a relation with target item $i$ to form $\mathcal{G}'_{II} = \{(j, r, i)|(j, r, i) \in \mathcal{G}, j \in \mathcal{I}_{u}, r \in \mathcal{R}_{II}\}$. c) We search an attribute $a \in \mathcal{A}$ from $\mathcal{G}_{IA}$ that associated to each item-item triplet in $\mathcal{G}'_{II}$, then connect this attribute to the items in each triplet to form $\mathcal{G}'_{IA}$. Eventually, we regard each triplet $(h, r, t)$ as a reason flow from $h$ to $t$, then identify and filter all triplet paths $\mathcal{P}_{R}$ that could reason from $u$ to $i$ within $\mathcal{G}'_{UI}$ and $\mathcal{G}'_{II}$.

\textbf{Path Analysis and Confidence-based Reasoning} To guide LLM in selecting the most suitable path $p$ from $\mathcal{P}_{R}$, we calculate the confidence score of each triplet in $\mathcal{G}'_{IA}$ to form a confidence set $\mathcal{C}_{IA}$. Confidence scores actually represent the global users' behaviors in the dataset domain, which provides LLM with a general user preference of $u$. To enrich user-specific information, we then randomly select items in $\mathcal{I}_{u}$ excluded from the items selected in reason paths $\mathcal{P}_{R}$ and add interactions of user $u$ and these selected items into $\mathcal{G}''_{UI}$. We finally combine information from $\mathcal{P}_{R}$, $\mathcal{C}_{IA}$ and $\mathcal{G}''_{UI}$ to prompt LLM to determine the most plausible reasoning path and generate an explanation. Details are shown at Appendix.

\vspace{-0.2cm}\section{Experiments}

\begin{table}[t]
\vspace{-0.4cm}
\centering
\caption{Statistics of datasets.}
\vspace{-0.4cm}
\label{tab:datasets-statistics}
\resizebox{0.45\textwidth}{!}{
\begin{tabular}{c|l|ccc}
\toprule
& Statistics & AmazonBook & Steam & Anime \\
\midrule
\multirow{4}{*}{IG} & Users & 13373 & 53533 & 18394\\
& Items & 37837 & 13232 & 10228\\
& Interactions & 520546 & 2,035,039 & 3,995,448\\
& Density & $1e^{-3}$ & $2.8e^{-3}$ & $2.1e^{-2}$ \\
\midrule
\multirow{3}{*}{TKG} & Entities/Attributes & 34902 & 6,632 & 3423\\
& Relations & 7 & 7 & 5\\
& IA Triplets & 58233 & 44773 & 33000\\
\midrule
\multirow{2}{*}{Pools Augment} & User View($\mathcal{P}^{add}_{\mathcal{U}}$, $\mathcal{P}^{del}_{\mathcal{U}}$) & 60745/17372 & 61220/16866 & 19386/13352\\
& Item View($\mathcal{P}^{add}_{\mathcal{I}}$, $\mathcal{P}^{del}_{\mathcal{I}}$) & 61340/12925 & 57435/3796 & 18840/987\\
\bottomrule
\end{tabular}
}
\vspace{-0.4cm}
\end{table}
Here,
we conduct extensive experiments to evaluate our CKG-LLMA model comparing with various baselines.

\vspace{-0.2cm}\subsection{Experimental Settings}

\textbf{Datasets} We utilize three diverse datasets for evaluation, Amazon-Book~\cite{amazonbook} for book recommendations, Steam~\cite{steam} for game recommendations and Anime~\cite{anime} for anime recommendations, to ensure a comprehensive and realistic evaluation. To construct knowledge graphs, we incorporate item-specific attributes (e.g., authors, publishers, categories for Amazon-Book) as additional entities and establish item-item relationships based on shared attributes. We first utilize the LLM-based Subgraph augmenter to query ChatGPT-turbo-3.5 with modification advice of origin KG, and we construct subgraphs with 32 collaborative signals for AmazonBook and 40 for Steam and 64 for Anime. Table~\ref{tab:datasets-statistics} summarizes the statistics of the Interaction Graph (IG) and constructed Temporal Knowledge Graph (TKG) for datasets.

\textbf{Evaluation} For each user $u$ in the IG dataset, we split their interacted items into train, validation, and test sets with an 8:1:1 ratio. We employ a full-ranking strategy for evaluation, treating all uninteracted items of each user $u$ as negative samples and interacted items in the validation or test set as positive samples. We assess top-$N$ recommendations using Recall@$N$ and NDCG@$N$, with $N=10$. Reported results are averaged across all users in the test set. Hyperparameters are tuned using the validation set.

\textbf{Baselines} We compare CKG-LLMA with previous recomendation systems in four categories: a) Traditional collaborative filtering (CF) models: \textit{LightGCN}~\cite{lightgcn} a simplified GCN-based model for representing users and items. \textit{SGL}~\cite{sgl} is a self-supervised CF model that constructs multiple graph views for contrastive learning in GNN-based recommendation. \textit{SimGCL}~\cite{simgcl} is a self-supervised CF model that applies uniform noise to embeddings for generating contrastive views. \textit{XSimGCL}~\cite{xsimgcl} is an extended self-supervised CF model that enhances SimGCL by applying uniform noise to specific intermediate layers during GCN propagation. b) GNN-based knowledge-aware models: \textit{KGCN}~\cite{kgcn} is a knowledge-aware GNN model that aggregates items' high-order information in GNN and user preferences for recommendation. \textit{KGAT}~\cite{kgat} is a knowledge-aware GNN model that integrates attentive aggregation into a collaborative KG for enhanced recommendation. \textit{KGIN}~\cite{kgin} is a knowledge-aware GNN model that incorporates user intention modeling and relational path-aware aggretaion for interpretable recommendations. c) Self-supervised knowledge-aware models: \textit{KGCL}~\cite{kgcl} is a self-supervised knowledge-aware model that introduces graph contrastive learning for KG. \textit{KGRec}~\cite{kgrec} is a SOTA self-supervised knowledge-aware model that devises a rationalization method to weigh knowledge connections in KG.  d) LLM-based models: \textit{LLMRec}~\cite{LLMRec} generates semantic embeddings for user and item embeddings and incorporates with LLMs. \textit{RLMRec}~\cite{rlmrec} is a SOTA LLM-based model which acquires text profiles from LLM and combine them with recommenders through contrastive learning.

\begin{table}[t]
\vspace{-0.4cm}
\centering
\caption{Performance comparisons. The best results are in bold, and the best baseline results are underlined.}
\vspace{-0.4cm}
\label{tab:model-comparison}
\resizebox{0.45\textwidth}{!}{
\begin{tabular}{ccccccc}
\toprule
\multirow{2}{*}{Model} & \multicolumn{2}{c}{AmazonBook} & \multicolumn{2}{c}{Steam} & \multicolumn{2}{c}{Anime} \\
\cmidrule(r){2-3} \cmidrule(l){4-5} \cmidrule(l){6-7}
 & Recall & NDCG & Recall & NDCG & Recall & NDCG \\
\midrule
LightGCN & 0.1583 & 0.1289 & 0.0601 & 0.0476 & 0.0631 & 0.1507 \\
SGL      & 0.1817 & 0.1459 & 0.0521 & 0.0474 & 0.0817 & 0.1659 \\
SimGCL   & 0.1627 & \underline{0.1519} & 0.0549 & 0.0444 & 0.0827 & 0.1719 \\
XSimGCL  & 0.1525 & 0.1330 & 0.0619 & 0.0460 & 0.0825 & 0.1815 \\
\midrule
KGCN     & 0.1657 & 0.1128 & 0.0475 & 0.0467 & 0.0757 & 0.1628 \\
KGAT     & 0.1676 & 0.1244 & 0.0429 & 0.0489 & 0.0776 & 0.1644 \\
KGIN     & 0.1625 & 0.1362 & 0.0524 & 0.0501 & 0.0825 & 0.1762 \\
\midrule
KGCL     & 0.1738 & 0.1311 & 0.0633 & 0.0508 & 0.0967 & 0.2290 \\
KGRec    & \underline{0.1821} & 0.1518 & \underline{0.0651} & 0.0521 & \underline{0.1010} & 0.2458 \\
\midrule	
RLMRec     & 0.1795 & 0.1402 & 0.0623 & 0.0498 & 0.0997 & \underline{0.2459}\\
LLMRec    & 0.1672 & 0.1420 & 0.0588 & \underline{0.0524} & 0.0982 & 0.2307\\
\midrule
CKG-LLMA & \textbf{0.1883} & \textbf{0.1592} & \textbf{0.0672} & \textbf{0.0546} & \textbf{0.1069} & \textbf{0.2518} \\
\bottomrule
\end{tabular}
}
\vspace{-0.33cm}
\end{table}

\begin{table}[t]
\centering
\caption{Ablation results with different variants.}
\vspace{-0.2cm}
\label{tab:ablation}
\resizebox{0.45\textwidth}{!}{
\begin{tabular}{ccccccc}
\toprule
\multirow{2}{*}{Settings} & \multicolumn{2}{c}{AmazonBook} & \multicolumn{2}{c}{Steam} & \multicolumn{2}{c}{Anime}\\
\cmidrule(r){2-3} \cmidrule(l){4-5} \cmidrule(l){6-7}
& Recall & NDCG & Recall &  NDCG & Recall & NDCG \\
\midrule
CKG-LLMA & \textbf{0.1883} & \textbf{0.1592} &  \textbf{0.0672} & \textbf{0.0546} & \textbf{0.1069} & \textbf{0.2518} \\
\midrule
w/o contrastive & 0.1312 & 0.1241 & 0.0508 & 0.0482 &  0.0743 & 0.1960  \\
w/o confidence  & 0.1773 & 0.1557 & 0.0622 & 0.0501 &  0.0920 & 0.2014  \\
20\% LLM        & 0.1842 & 0.1531 & 0.0623 & 0.0502 &  0.1026 & 0.2156   \\
40\% LLM        & 0.1857 & 0.1547 & 0.0649 & 0.0525 &  0.1037 & 0.2234   \\
60\% LLM        & 0.1868 & 0.1560 & 0.0660 & 0.0537 &  0.1057 & 0.2331   \\
80\% LLM        & 0.1872 & 0.1569 & 0.0668 & 0.0540 &  0.1062 & 0.2477   \\
\bottomrule
\end{tabular}
}
\vspace{-0.4cm}
\end{table}

\vspace{-0.2cm}\subsection{Overall Performance Comparison}
We report the performance of all the methods on three datasets in Table \ref{tab:model-comparison}. There are some observations: 

\textbf{Overall Model Comparison}: Our model CKG-LLMA consistently performs better than other baselines in the majority of cases, reflecting the effectiveness of LLM augmented knowledge graphs and the designed confidence mechanism. The consistent improvement across various metrics demonstrates the robustness and generalizability of our model.

\textbf{Impact of Confidence Mechanism through Knowledge-aware Methods}: Recent knowledge-aware models show improved recommendation performance compared to traditional CF methods, highlighting the potential of incorporating external knowledge into recommendation systems. However, a notable performance disparity persists between some KG-based approaches and self-supervised learning CF methods, such as XSimGCL. We attribute this discrepancy to the proliferation of complex item attributes within KGs that contribute minimally to recommendation tasks. In contrast, our model adopts a confidence-based probability approach to refine triplets, ensuring that only useful knowledge contributes to the recommendation process.

\textbf{Merits Compared to Graph Contrastive Learning Approaches}: Both traditional and knowledge-aware models with graph contrastive learning endeavors significantly outperform other baselines, underscoring the efficacy of contrastive loss in enhancing the quality of modeling latent representations for user preferance. Among these, our model demonstrates superior performance. Unlike other models that often rely on random graph augmentation or intuitive handcrafted cross-view pairing to build contrastive views, our model leverages the power of LLMs to infuse KG with richer semantic information and employs a confidence-aware two-step augmentation mechanism to construct contrastive objectives that are closely aligned with the inherent nature of recommendation systems. 

\vspace{-0.2cm}\subsection{Ablation Study}
In the ablation study, we assess the individual contributions of CKG-LLMA with the following variants. The results are shown in table \ref{tab:ablation}.

\textbf{w/o Confidence}: We disable the confidence-based dropout process in knowledge graph augmentation. The observed significant performance degradation underscores the efficacy of the confidence-aware differentiable KG augmentation in capturing valuable semantics generated by LLM to enhance the KG structural information.

\textbf{w/o Contrastive Loss}: We nullified the contrastive loss by setting its weight $\lambda_c$ to zero, training the model solely on $l_{bpr}$ without LLM-enhanced information or the confidence-aware denoising mechanism. The resultant performance degradation indicates that user/item representations become contaminated with KG noise in the absence of contrastive learning, significantly impacting model efficacy.

\textbf{Different Proportions of LLM}: To assess the quality of LLM-generated information, we randomly sampled triplets from $\mathcal{P}^{add}_{\mathcal{U}}$, $\mathcal{P}^{del}_{\mathcal{U}}$, $\mathcal{P}^{add}_{\mathcal{I}}$, $\mathcal{P}^{del}_{\mathcal{I}}$ at ratios of 20\%, 40\%, 60\%, and 80\%. We then evaluated our model's performance using these subsets, maintaining consistent add ratio and delete ratio with the 100\% baseline. The observed positive correlation between performance metrics and sampling ratios suggests that the LLM-based Subgraph Augmenter effectively identifies latent entity relationships within KG, thereby enriches auxiliary information and enhances user modeling.


\begin{figure}[t]
\vspace{-0.3cm}
\centering
\begin{subfigure}[t]{0.45\textwidth}
  \centering
  \includegraphics[width=\textwidth]{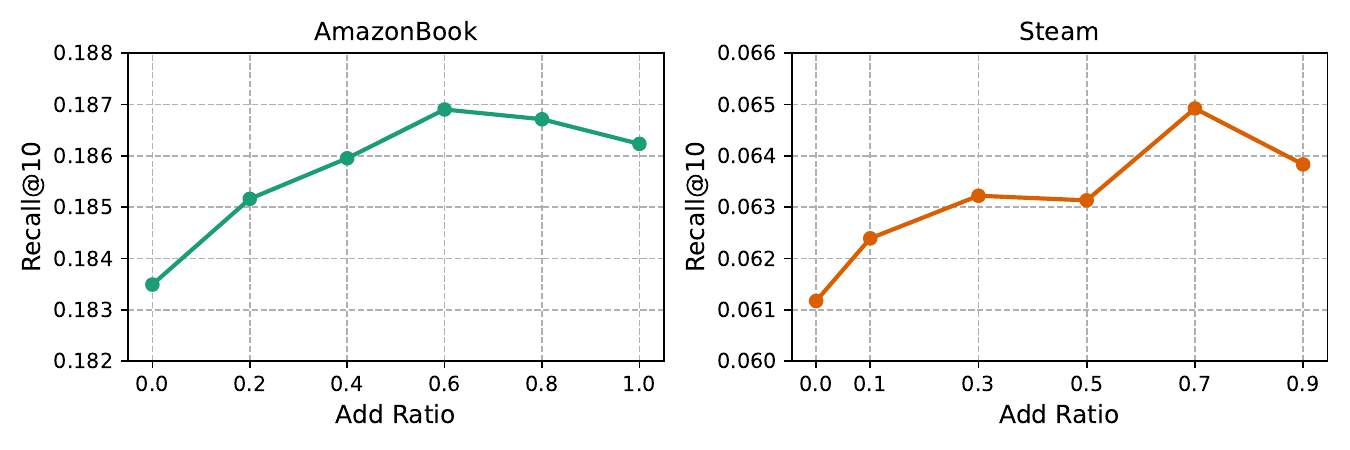}
  \label{fig:add_ratio_pa}
\end{subfigure}
\begin{subfigure}[t]{0.45\textwidth}
\vspace{-0.4cm} 
  \centering
  \includegraphics[width=\textwidth]{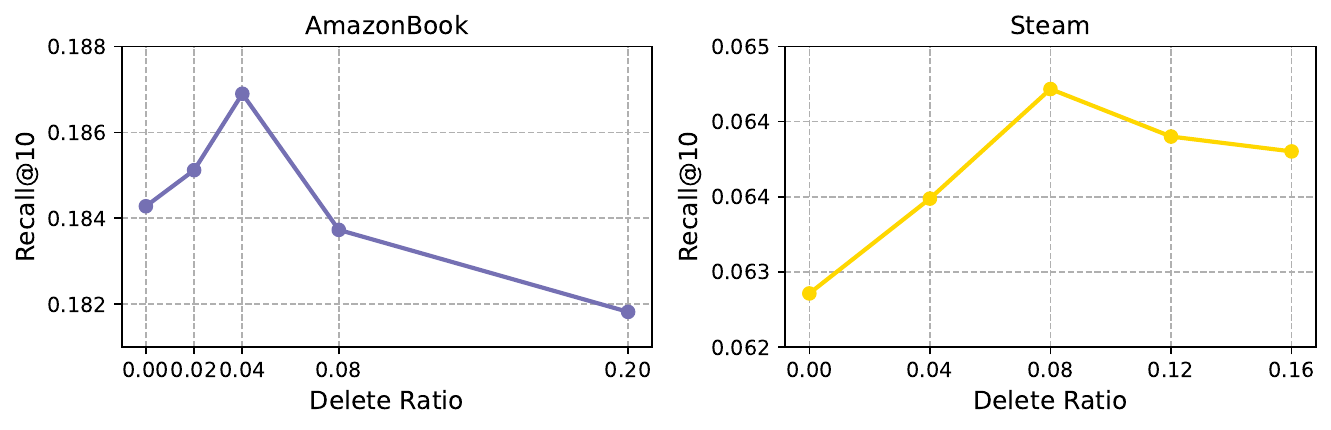}
  \label{fig:delete_ratio_pa}
\end{subfigure}
\vspace{-0.6cm}
\caption{Parameter Analysis of add ratio and delete ratio.}
\label{fig:combined_ratio_pa}
\end{figure}

\vspace{-0.3cm}
\subsection{Parameter Analysis}
We analyze our model's sensitivity to the following parameters: \textbf{Delete ratio $\mu_d$} and \textbf{Add ratio $\mu_a$}. The results are presented in Figure \ref{fig:combined_ratio_pa}. We varied the add ratio from 0\% to 100\% across two datasets while fixing $\mu_d$ at 4\% for AmazonBook and 8\% for Steam. The results show that CKG-LLMA achieves best performance at $\mu_a = 60\%$ for the AmazonBook dataset and $\mu_a = 70\%$ for the Steam dataset. As $\mu_a $ increases, more information is extracted from the generated pools in each step, and our model demonstrates a robust capability to process varied amount of information, affirming the efficacy of our confidence mechanism. Notably, despite the increase in $\mu_a$, the model's overall performance remains relatively stable, exhibiting only minor fluctuations, which indicates its robust noise-resistance capabilities. Due to the different sizes of the generated deletion pools $\mathcal{P}^{del}$ in the two datasets, we varied the delete ratio from 0\% to 20\% in the AmazonBook dataset with $\mu_a$ fixed at 60\%, and from 0\% to 16\% in the Steam dataset with $\mu_a$ fixed at 70\%. The results indicate that CKG-LLMA achieves optimal performance around $\mu_d = 4\%$ for the AmazonBook dataset and $\mu_d = 8\%$ for the Steam dataset.Moreover, the differences between various variants and our method have been statistically validated through significance hypothesis testing, further underscoring the robustness of our approach.

\begin{table}[t]
\centering
\vspace{-0.3cm} 
\caption{Case study with the Confidence-aware Explanation Generation procedure.}
\vspace{-0.3cm} 
\label{tab:casestudy}
\resizebox{0.45\textwidth}{!}{
\begin{tabular}{lp{10cm}}
\toprule\toprule
User-Item pair & (\textit{user8614}, \textit{game4705}) in Steam dataset\\
User's review & You can tell the \textbf{devs} put a lot of love and hard work into this game. \\
\midrule
\multirow{3}{*}{Candidate Paths} & $user8614 \rightarrow{}  game2949  \xleftrightarrow[\text{Single Player}]{\text{Same spec}} game4705 \left[2.132, 2.561\right]$ \\
 & $user8614 \rightarrow game1858 \xleftrightarrow[\text{Devolver Digital}]{\text{Same publisher}} game4705 \left[2.627, 2.773\right]$
 \\
 & $user8614 \rightarrow{}  game7489  \xleftrightarrow[\text{Action}]{\text{Same genre}} game4705\left[1.528, 1.334\right]$ \\
LLM select & $user8614 \rightarrow{}  game1858  \xrightarrow{\text{same publisher}} game4705$ \\
Ours(with confidence) & \textbf{Devolver Digital} delivers another gem. Shadow Warrior 2 maintains their high standards and unique style. \\
Ours(w/o confidence) & Shadow Warrior 2's single-player experience lets you immerse in the game world at your own pace.\\
\bottomrule\bottomrule
\end{tabular}
}\vspace{-0.4cm} 
\end{table}

\vspace{-0.2cm}\subsection{Case Study in Explanation Generation}
Here we present case to show the explainable power of our model. We select a positive pair from the testset of Steam dataset, (\textit{user8614}, \textit{game4705}), where the corresponding triplet (\textit{user8614, interacted, game4705}) is not in augmented $\mathcal{G}^{aug}$. Details are shown in table \ref{tab:casestudy}. We first collect all items which has been interacted with \textit{user8614}, and filter out the rest items (\textit{game2949}, \textit{game1858}, \textit{game7489}) which has an item-item relationship with target item \textit{game4705}. We construct the candidate paths based on the 3 items, and calculate confidence scores of triplets in each path (eg. Confidence of (\textit{game2949}, \textit{has spec}, \textit{Single Player}) is 2.132). We then query LLM with a prompt combining interacted items of \textit{user8614}, candidate paths and associated confidences scores and derive the user-review alike explanation. The result indicates that Confidence scores enable LLM to better understand path relevance to user preferences, while without them, LLM resort to random path selection.

\vspace{-0.2cm}\section{Conclusion}
In this paper, we introduced the Confidence-aware KG-based Recommendation Framework with LLM Augmentation (CKG-LLMA), which effectively integrates Large Language Models (LLMs) with Knowledge Graphs (KGs) to enhance recommendations. Our framework introduces an LLM-based subgraph augmenter for enriching KGs, implements a confidence-aware mechanism for filtering noise, and employs a dual-view contrastive learning schema to enhance user-item representations. Experiments show that CKG-LLMA outperforms existing methods and generates high-quality, interpretable recommendations.

\bibliographystyle{named}
\bibliography{ijcai25}

\clearpage
\begin{figure*}
    \centering
    \includegraphics[width=0.9\linewidth]{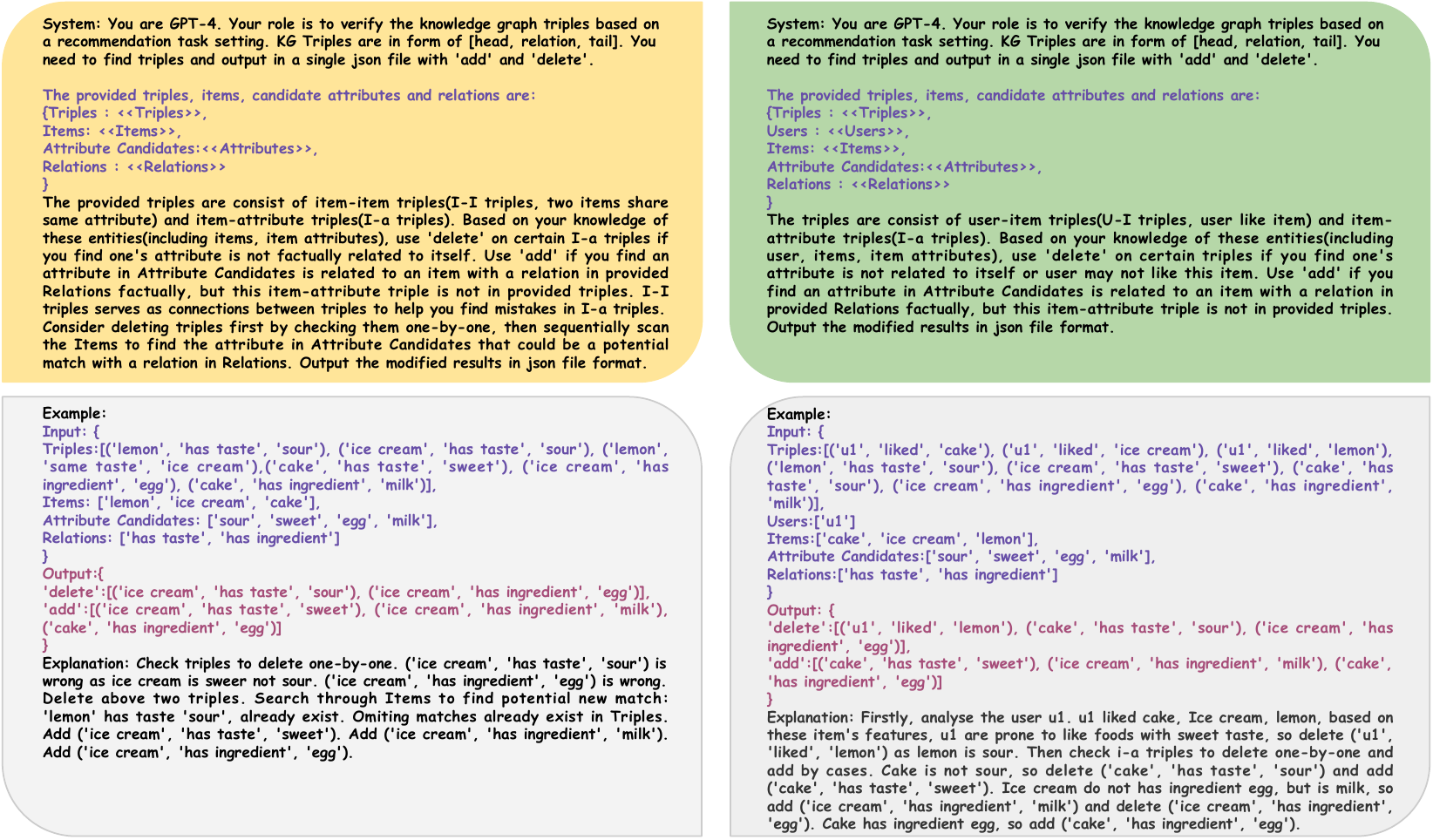}
    \caption{Prompts for LLM-based Subgraph Augmenter. Upper left is Item-view prompt and lower left is its related example. Upper right is User-view prompt and lower right is its related example.}
    \label{fig:subgraph_prompt}
\end{figure*}

\begin{figure*}
    \centering
    \includegraphics[width=0.9\linewidth]{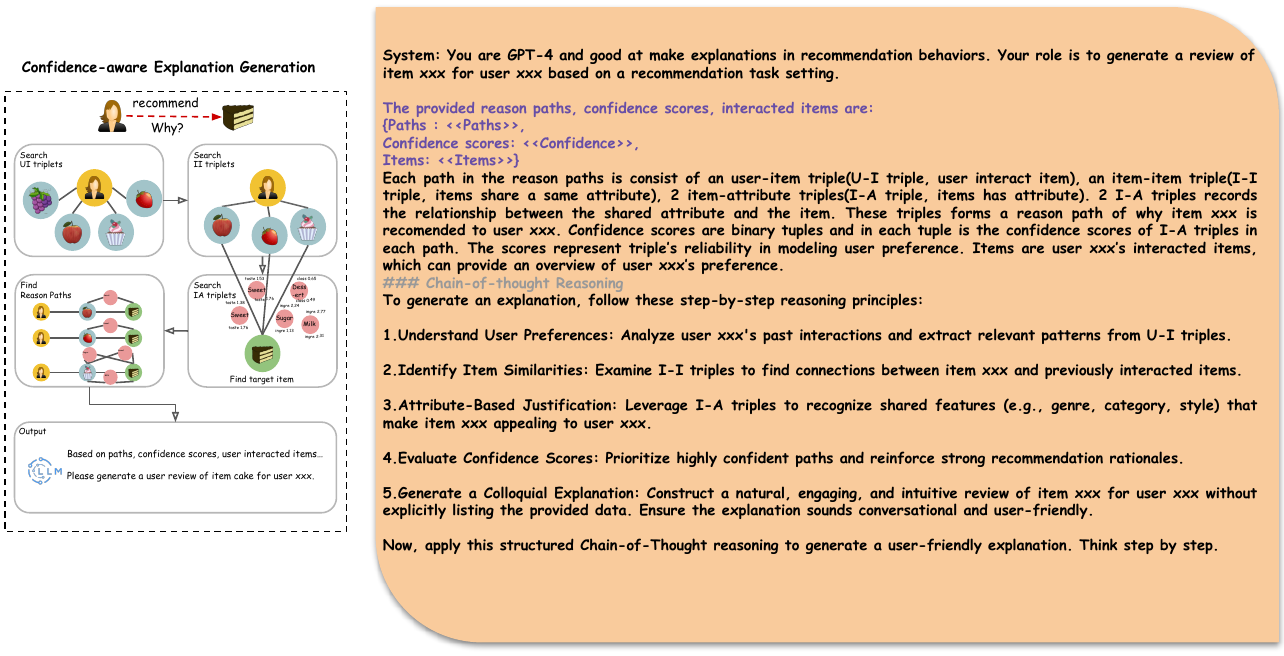}
    \caption{The overall pipeline for Confidence-aware Explanation Generation. The left is the framework of generation process, and the right is the prompt template of LLM generation.}
    \label{fig:explanation-generation-pip}
\end{figure*}

\section{Appendix}

\subsection{Prompt Details}

\subsubsection{Prompts in LLM-based Subgraph Augmenter}

The designed prompts for augmenting subgraphs in two views are shown in figure~\ref{fig:subgraph_prompt}. The differences between prompts of two views are mainly the provided information, the constitution of triples and the designed tasks, in which User-view prompt instructs LLM to rectify I-A triplets through fact verification, and prune the provided user-item interactions through reasoning; Item-view prompt directs LLM to meticulously rectify I-A triplets and utilizes auxiliary information from I-I triplets to enhance the rectification process.

\subsubsection{Pipeline and Prompt in Confidence-aware Explanation Generation}

The overall pipeline of the explanation generation module and the designed prompt template are shown in figure~\ref{fig:explanation-generation-pip}. For certain user-item interaction pair ($u$, $i$), the module initially retrieves interacted items of user $u$ to search for relevant U-I triplets. Then we identify items with an item-item relationship to target item $i$ and derive all corresponding I-I triplets. In the next step, find the co-occupied attribute in each I-I triplet and obtain two item-entity triplets based on this attribute. We then categorizes collected triplets into distinct reason paths, and explicates each component and the structure of these paths in the prompt. Eventually, we fill the prompt template with collected information to instruct LLM to generate an review-alike explanation from user $u$ to item $i$.

The reasoning process follows a structured chain-of-thought approach to ensure a coherent and interpretable explanation. To achieve this, we apply the following step-by-step approach based on \cite{cot}:
\begin{itemize}
    \item 1. \textbf{Understand User Preferences}:Analyze user $u$'s past interactions and extract relevant patterns from U-I triples, providing an overview of the user's previous choices.
    \item2. \textbf{Identify Item Similarities}: Examine I-I triples to find connections between item $i$ and previously interacted items, emphasizing the contextual relevance.
    \item3. \textbf{Attribute-Based Justification}: Leverage I-A triples to recognize shared features (e.g., genre, category, style) that make item $i$ appealing to user $u$.
    \item4. \textbf{Evaluate Confidence Scores}: Prioritize highly confident paths and reinforce strong recommendation rationales by assigning reliability scores to I-A triples.
    \item5. \textbf{Generate a Colloquial Explanation}: Construct a natural, engaging, and intuitive review of item $i$ for user $u$ without explicitly listing the provided data, ensuring that the explanation remains user-friendly and comprehensible.
\end{itemize}

By following these structured reasoning steps, the LLM is guided to generate coherent, persuasive, and contextually relevant explanations that align with user preferences and in

\subsection{Hyperparameter Configuration}

The hyperparameter setting of CKG-LLMA in dataset AmazonBook is shown at table~\ref{tab:hypersettings}.

\begin{table}[t]
\centering
\caption{Hyperparameter settings of CKG-LLMA in AmazonBook dataset}
\label{tab:hypersettings}
\begin{tabular}{c|c}
\toprule
Hyperparameter & Settings \\
\midrule
Learning rate $l_r$ & 1e-4 \\
Hidden dimension $d$ & 64 \\
Subgraph size $N$ & 32 \\
Delete ratio $\mu_d$ & 0.08 \\
Add ratio $\mu_a$ & 0.60\\
Num of Experts $N_e$ & 8 \\
Scalable parameter $K$ & 5 \\
Temperature in gumbel softmax $\tau_{g}$ & 0.9 \\
Drop probability $p_{drop}$ & 0.01 \\
Message propogation layers $L$ & 3\\
Temperature in contrastive loss $\tau$ & 0.2 \\
contrastive loss weight $\lambda_{\theta}$ & 1e-3 \\
regularization loss weight $\lambda_{\theta}$ & 1e-4 \\
\bottomrule
\end{tabular}
\end{table}

\section{Model Complexity Analysis}
CKG-LLMA maintains comparable parameter and time complexity to SOTA models like LLM-based~\cite{rlmrec}, KG-based~\cite{kgcl} works. Regarding model complexity(W1), the learnable parameters include user, item, and attribute embeddings, with a complexity of $\mathcal{O}((|U|+|I|+|A|)\cdot d)$, plus a confidence-aware aggregation with $\mathcal{O}(2d^2)$. Notably, Our model maintains comparable parameter complexity with KGCL, while being more efficient than KGRec through fewer modules and loss functions, and achieves better parameter efficiency than alignment-based LLMRec and RLMRec. Time complexity scales linearly with the number of user-item interactions and KG triplets in dataset, consisting these components: (1)Confidence-aware propagation with $\mathcal{O}(|G_{IA}|\cdot|R|\cdot d)$ for confidence computation and $\mathcal{O}(|G_{IA}|\cdot d)$ for aggregation. (2)Differentiable KG augmentation with $\mathcal{O}(|G_{IA}| \cdot (|R|\cdot d+1))$. (3)Contrastive learning with LightGCN $\mathcal{O}(|G_{UI}|\cdot d)$ and con-loss $\mathcal{O}(B_s\cdot(|U|+|I|)\cdot d)$. 

\section{Experimental Cost}
The cost is low as we only uses
LLMs during preprocessing. For example, the CHATGPT API costs just 27usd on AmazonBook
and 68usd on Steam, as CKG-LLMA performs subgraph augmentation offline,
with no LLM inference during training. While we used GPT-3.5-turbo for its performance and
affordability, CKG-LLMA can also use open-source LLMs. 
\end{document}